\documentclass[%
reprint,
 amsmath,amssymb,
 aps,
floatfix,
]{revtex4-1}

\usepackage{color}
\usepackage{graphicx}
\usepackage{bm}
\usepackage{hyperref}

\begin{document}

\title{Spin dependence of the tricritical point in the mixed-spin Blume-Capel model on three-dimensional lattices: Metropolis and Wang-Landau sampling approaches}

\author{Mouhcine Azhari}
\email{azhari.mouhcine@gmail.com}
\affiliation{Laboratory of High Energy Physics and Condensed Matter, Hassan II University-Casablanca, Faculty of Sciences A\"{\i}n-Chock, B.P: 5366 Maarif, Casablanca 20100, Morocco}

\author{Unjong Yu}
\email{Corresponding author: uyu@gist.ac.kr}
\affiliation{Department of Physics and Photon Science, Gwangju Institute of Science and Technology, Gwangju 61005, South Korea}

\date{\today}

\begin{abstract}
We investigate the mixed-spin Blume-Capel model with spin-1/2 and spin-$S$ ($S=1$, $2$, and $3$) on the simple cubic and body-centered cubic lattices with single-ion-splitting crystal-field ($\Delta$) by using the Metropolis and the Wang-Landau Monte Carlo methods. 
By numerical simulations, we prove that the tricritical point is spin-independent for both lattices. 
The positions of the tricritical point in the phase diagram are determined as ($\Delta_t/J=2.978(1)$; $k_B T_t/J=0.439(1)$) and ($\Delta_t/J=3.949(1)$; $k_B T_t/J=0.854(1)$) for the simple cubic and the body-centered cubic lattices, respectively. 
A very strong supercritical slowing down and hysteresis were observed in the Metropolis update close to first-order transitions for $\Delta>\Delta_t$.
In addition, for both lattices we found a line of compensation points, where the two sublattice magnetizations have the same magnitude.
We show that the compensation lines are also spin-independent.
\end{abstract}

\maketitle


\section{Introduction}

The Ising model \cite{Ising} is one of the simplest and the most studied cooperative many-body models in the communities of statistical mechanics and condensed matter physics that can be solved analytically on one- and two-dimensional lattices \cite{Onsager}.
Despite the tremendous effort in the last few decades, there is unfortunately no analytic solution in three dimensions (for a recent review, see Ref.~\cite{Ferrenberg18}). 
Nevertheless, this model is an indispensable tool for answering scientific questions in diverse research areas. It has been used for various physical systems, such as a model for certain kinds of highly anisotropic magnetic crystals as well as a lattice model for fluids, alloys, adsorbed monolayers and even more in field theories of elementary particles \cite{Enting79}. 
It was also used successfully for biological and chemical systems, and in the design of quantum computers based on one-dimensional Ising systems \cite{Leuenberger01,Berman94}.

Interesting features may arise when one considers more than two states and the Blume-Capel model \cite{Blume66,Capel66} is one of the simplest extensions.
This model has attracted particular attention in connection with its wetting and interfacial adsorption under the presence or absence of bond randomness~\cite{Selke83,Selke84,Fytas13,Fytas19}.
It consists of a spin-1 Ising Hamiltonian with an anisotropy field (also called single-ion-splitting crystal-field).
The latter term controls the density of vacancies and plays a dominant role in the existence of the tricriticality. 
It allows the model to have a tricritical point (TCP) in two- and three-dimensional lattices \cite{Beale86,Deserno97,Silva06,Fytas11,Kwak15,Jung17,Butera18}. 
However, this situation changes radically when we consider a mixture of spin-1/2 and spin-$S$, since they have less translational symmetry than their single spin counterparts. This latter property has a great influence on the magnetic properties of the mixed-spin systems and causes them to exhibit unusual behavior not observed in single-spin Ising models. 
These mixed-spin models have already found various applications for the description of certain types of ferrimagnetism, such as the MnNi(EDTA)-6H$_2$O complex and the two-dimensional compounds A$^\mathrm{I}$M$^\mathrm{II}$Fe$^\mathrm{III}$(C$_2$O$_4$)$_3$ (A = N($n$-C$_3$H$_7$)$_4$,
N($n$-C$_4$H$_9$)$_4$, N($n$-C$_5$H$_{11}$)$_4$, P($n$-C$_4$H$_9$)$_4$, P(C$_6$H$_5$)$_4$, N($n$-C$_4$H$_9$)$_3$(C$_6$H$_5$CH$_2$), (C$_6$H$_5$)$_3$PNP(C$_6$H$_5$)$_3$, As(C$_6$H$_5$)$_4$; M$^\mathrm{II}$ = Mn, Fe) \cite{Drillion83,Mathoniere96}. 

Up to now, the mixed-spin Blume-Capel model has been explored in two varieties of exact analytical approaches in two dimensions \cite{Gon_alves85,Dakhama98,Dakhama18}. 
The first one is to use exact mapping transformation, which maps the subject model onto an exactly solved one with effective mapped interaction. 
Based on this transformation, the mixed spin-1/2 and spin-$S$ ($S>1/2$) Blume-Capel models on the honeycomb and Lieb lattices were solved exactly \cite{Gon_alves85,Dakhama98}. 
Moreover, one of us (M A) proposed recently a heuristic exact approach on the basis of a conjecture to solve the same models on the square lattice \cite{Dakhama18}. 
As a result, it turned out that in two-dimensional lattices the mixed-spin Blume-Capel model undergoes a continuous transition for all values of the crystal-field interactions for any half-integer $S>1/2$.
For integer $S$, long-range ordering disappears for crystal-field larger than a critical value.
The results of the Monte Carlo (MC) simulations \cite{Zhang93,Buendia97,Buendia99,Selke10} and the renormalization-group method \cite{Benayad90} are in full agreement with the analytic results  \cite{Gon_alves85,Dakhama98,Dakhama18}.

On the contrary, there is no exact result in three-dimensional cases. 
Selke and Oitmaa \cite{Selke10} have performed MC simulations for the Blume-Capel model with mixed spin-1/2 and spin-1 on the simple cubic (SC) lattice applying the Metropolis update (MU) of single-spin flips \cite{Metropolis53} and long runs.
They found evidence for a tricritical point (TCP) and a line of compensation points. 
This is consistent with the renormalization-group calculations \cite{Quadros94}, which indicates the existence of the TCP in the phase diagram. 
In Ref.~\cite{Selke10}, the location of the TCP was obtained to be $2\Delta_t/J = 5.91 \pm 0.03$ tentatively based on the histogram of the magnetization in a small system (linear size $L=4$ in our convention described in Sec.~\ref{Sec:Method}).
An accuratedetermination of the location of the TCP was beyond the scope of Ref.~\cite{Selke10}.
We show in this paper that such a small lattice size leads to an underestimation of the value $\Delta_t$ of the TCP, which calls for a more extensive study on this model


Motivated by this, we reexamine the mixed-spin Blume-Capel model on the SC lattice using the standard MU \cite{Metropolis53} and the Wang-Landau (WL) algorithm \cite{Wang01,Silva06,Fytas11,Kwak15}. As far as we know, the WL algorithm has never been implemented to study mixed-spin systems. We show that the two methods are complementary to each other. We propose a reliable method to locate the TCP for the mixed spin-1/2 and spin-1 Blume-Capel model on the SC lattice. The cases of $S=2$ and $3$ are also studied, and the spin-dependence of the TCP is discussed. The same method is applied to the body-centered cubic (BCC) lattice to locate the TCP for integer $S$. In addition, we show that both lattices exhibit compensation phenomena, which can be very useful in magnetic memory and spin analyzing applications \cite{Kumar15}. Our purpose here is to study these two lattices to better improve our understanding of the mixed-spin systems.

The outline of the article is as follows. As necessary background, in Sec.~\ref{Sec:Method} we introduce the mixed-spin Blume-Capel model and summarize the numerical details of our simulations. In Sec.~\ref{Sec:Results} we discuss our results for both lattices SC and BCC, and we present our analysis for the cases $S=1$, $2$, and $3$. We then conclude with a summary in Sec.~\ref{Sec:Conclusion}.

\begin{figure*}
\includegraphics[width=1.8\columnwidth]{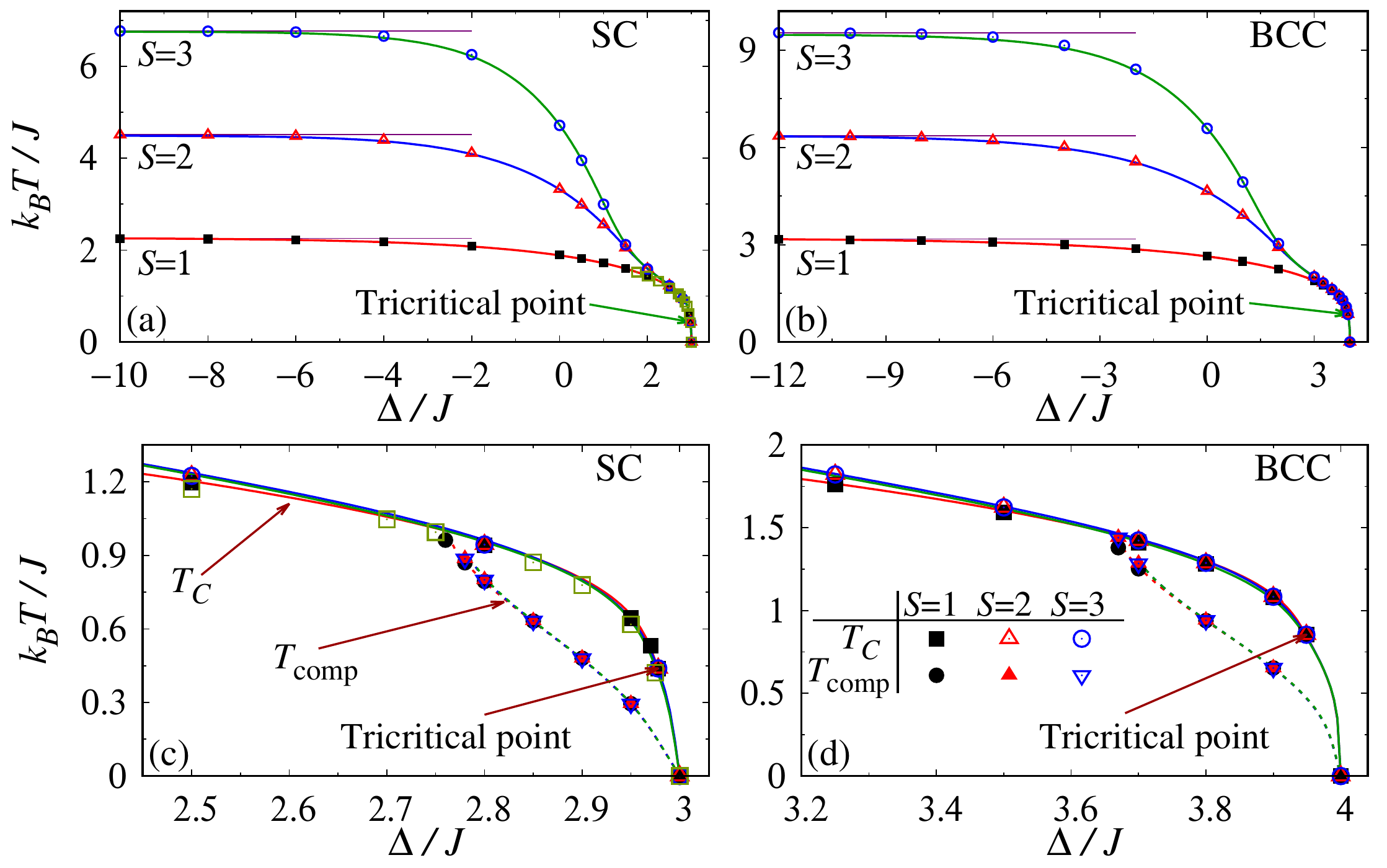}
\caption{\label{fig:fig1} 
Phase diagram in the $\Delta$-$T$ plane for the SC ((a) and (c)) and BCC ((b) and (d)) lattices. Compensation temperature ($T_\mathrm{comp}$) is represented with dashed lines in lower panels. The horizontal straight lines represent $T = (S/2)T_C^{\mathrm{Ising}}$, which is the critical temperature expected in the limit $\Delta \rightarrow - \infty$. $T_C^{\mathrm{Ising}}$ is the critical temperature of the conventional Ising model in the SC and BCC lattices. The transition temperature ($T_C$) and the compensation temperature ($T_\mathrm{comp}$) were obtained by using the WL (solid and dashed lines) and MU (symbols). The statistical error is smaller than the symbol size. Previous results of the transition temperature by Selke and Oitmaa \protect\cite{Selke10} in the SC lattice are also shown by the empty squares.}
\end{figure*}

\section{Model and Methods \label{Sec:Method}}

We studied the mixed-spin Blume-Capel model on the SC and BCC lattices. 
The Hamiltonian can be written as 
\begin{eqnarray}
H=-J \sum_{\langle i \in \Lambda_1, j \in \Lambda_2 \rangle} \sigma_i S_j + \Delta \sum_{j \in \Lambda_2} \left( S_j \right)^2 . \label{Hamiltonian}
\end{eqnarray}
Each lattice consists of two interpenetrating sublattices $\Lambda_1$ with the spin variables $\sigma_i$ and $\Lambda_2$ with spins $S_j$.
Spins $\sigma_i$ and $S_j$ may take on the values $\pm 1/2$ and $\{-S,-S+1,\cdots,S\}$, respectively, where $S$ is an integer or half-integer greater than $1/2$. 
In this paper, we study only integer $S$ cases ($S=1$, $2$, and $3$).
The notation $\langle i \in \Lambda_1, j \in \Lambda_2 \rangle$ stands for summation over all pairs of nearest-neighbor spins. 
The exchange interaction $J$ is between two nearest neighbors $\sigma_i$ and $S_j$, and $\Delta$ is the single-spin anisotropy. 
Positive $J$ means that the interaction is ferromagnetic. 
Since the lattices we study in this paper are bipartite, ferrimagnetic case ($J<0$) is completely equivalent to the ferromagnetic case. 
In this work, all the results presented in this paper are obtained for $J>0$.

We consider three-dimensional cubic lattices SC and BCC with the number of lattice points $N=BL^3$, where $L$ is the linear size of the system. 
The number of sites per unit cell $B$ takes the values 1 and 2 for the SC and BCC lattices, respectively. 
The periodic boundary condition is used in all directions. 
We used two kinds of MC schemes: the MU \cite{Metropolis53} and WL sampling \cite{Silva06,Fytas11,Kwak15}. 
The MU is simple, easy to implement, and provides access to simulations in large lattice sizes;
but it suffers from the critical and supercritical slowing down \cite{Janke94} and it is not reliable close to first-order transitions.
In contrast, the WL sampling overcomes the critical and supercritical slowing down and eliminates hysteresis. 
Besides, physical quantities for any temperature and anisotropy can be obtained just by one calculation. 
But the lattice size is limited in the WL method due to the multiparametric Hamiltonian of our model and hence the huge number of the energy levels, which increases with $S$. 
The maximum lattice size studied in this work is $L=100$ and $L=10$ for the MU and WL methods, respectively.
The MU has been widely used in both single and mixed-spin systems and we shall only discuss the relatively new method, the WL sampling. 

The WL sampling method directly estimates the density of states $\rho(E_1, E_2)$ via a random walk in energy space with the transition probability 
\begin{eqnarray}
P[(i_1, i_2)\rightarrow (j_1, j_2)] = 
   \mbox{min}\left[1, \frac{\rho(E_{i_1}, E_{i_2})}{\rho(E_{j_1}, E_{j_2})}\right],
\end{eqnarray}
which makes histogram $h(E_1, E_2)$ flat.
The two energy variables $E_1$ and $E_2$ represent the two terms of the Hamiltonian in Eq.~(\ref{Hamiltonian}), respectively:
\begin{eqnarray}
E_1 = \sum_{\langle i \in \Lambda_1, j \in \Lambda_2 \rangle} \sigma_i S_j ~ 
\mbox{ and } ~
E_2 = \sum_{j \in \Lambda_2} \left( S_j \right)^2 .
\end{eqnarray}
The energy space $\xi$ of the density of states $\rho(E_1, E_2)$ is proportional to the size of the system $L$ and the spin variables $S$ as $\xi=(zSN/2)(S^2N/2)=(zB^2/4)L^6S^3$, where $z$ is the coordination number; about half of  $\xi$ has nonzero density of states.
We found that the CPU time required to get the density of states is roughly proportional to $\xi^{\alpha}$ with $\alpha=1.26(4)$; it takes about eight hours for $L=10$ and $S=1$ in the SC lattice on a 2.2 GHz Intel(R) Xeon(R) processor.

At each step, the WL refinement is $\rho(E_{1}, E_{2}) \rightarrow f_n \, \rho(E_{1}, E_{2})$, where $f_n>1$ is an empirical factor. 
Whenever the energy histogram is flat enough, the modification factor $f_n$ is adjusted as $f_{n+1}= \sqrt{f_n}$ with $f_0 = e$ and a new set of random walks is performed.
The whole simulation is terminated when $f_n$ becomes close enough to 1: $f_\mathrm{final} < \exp(10^{-10})$. See Ref.~\cite{Yu15} for more detail.
During the simulation, average values of thermodynamic observables $O(E_{1}, E_{2})$ as a function of $E_1$ and $E_2$ should be calculated.

Once the density of states $\rho(E_{i_1}, E_{i_2})$ is obtained, the partition function can be calculated for any values of temperature and anisotropy,
\begin{eqnarray}
Z(T,\Delta)=\sum_{E_1, E_2} \rho(E_1, E_2)  e^{\beta(J E_1 - \Delta E_2 )},
\end{eqnarray}
where $\beta$ denotes the inverse temperature $1/k_B T$ and $k_B$ is the Boltzmann constant. 
It is straightforward that all thermodynamic observables $\langle O \rangle(T,\Delta)$ can be calculated without additional simulation for each temperature and anisotropy:
\begin{eqnarray}
\langle O \rangle (T,\Delta)= \frac{1}{Z} \sum_{E_1, E_2} O(E_{1}, E_{2}) \rho(E_1, E_2) e^{\beta(J E_1 - \Delta E_2 )} . \nonumber
\end{eqnarray}
To map the phase diagram, we calculated the sublattice and the total magnetizations
\begin{eqnarray}
M_{\sigma} &=& \frac{ \left|\sum_{i \in \Lambda_1} \sigma_i \right| }{N_{\sigma}} , \\
M_{S} &=& \frac{ \left| \sum_{j \in \Lambda_2} S_j \right| }{N_{S}} ,\\
M &=& \frac{ \left|\sum_{i \in \Lambda_1} \sigma_i + \sum_{j \in \Lambda_2} S_j \right| }{N} .
\end{eqnarray}
Note that $N_{\sigma}=N_{S}=N/2$ for the SC and BCC lattices.
In addition, to locate the critical temperature $T_C$ and to determine the type of transition, we calculated the Binder cumulant \cite{Binder81}
\begin{eqnarray}
U= 1 - \frac{\left\langle M^4 \right\rangle}{3 \left\langle M^2 \right\rangle ^2} .
\end{eqnarray}

\begin{figure*}
\includegraphics[width=1.82\columnwidth]{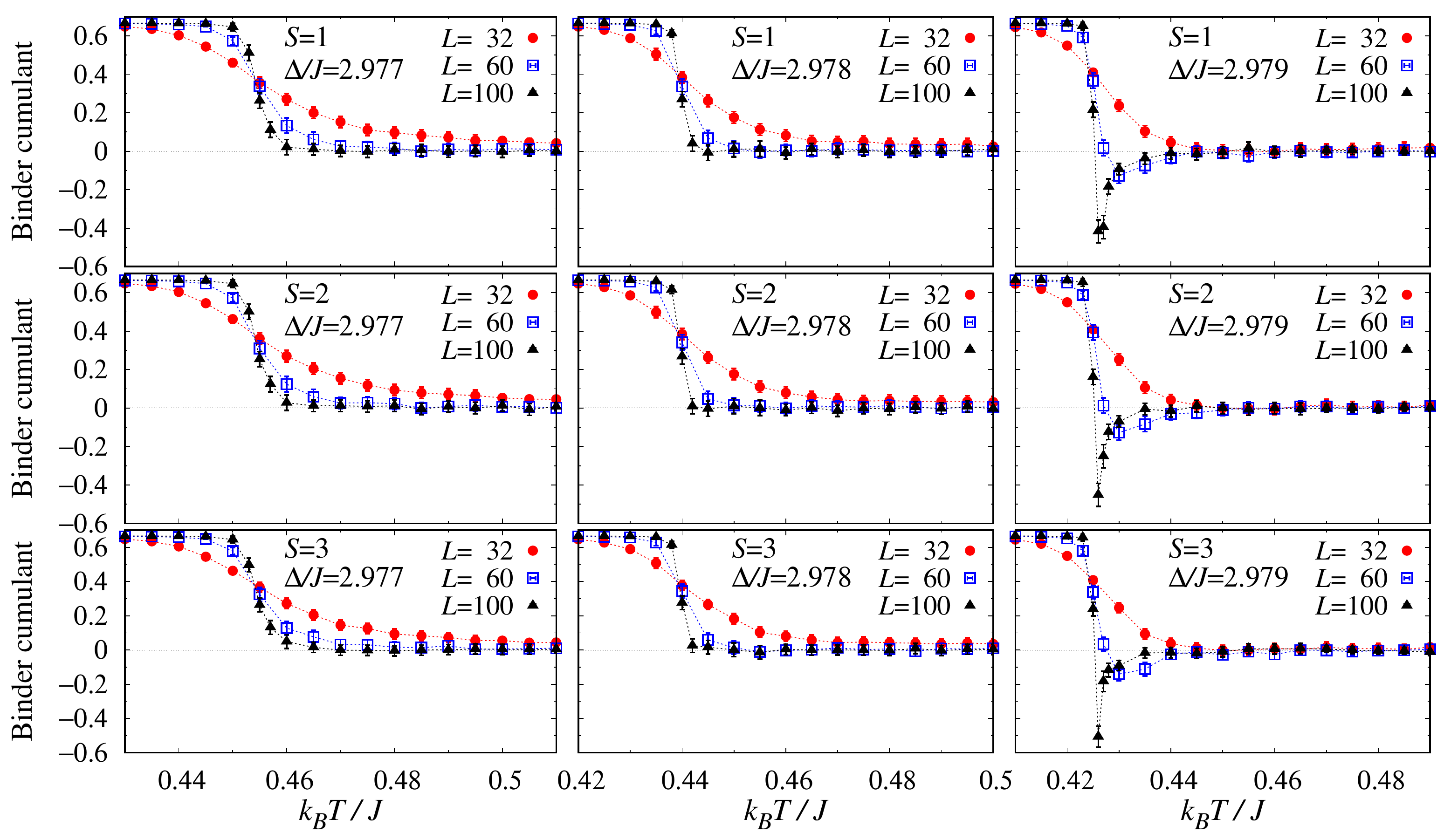}
\caption{\label{fig:fig2}Binder cumulant as a function of temperature ($T$) for various lattice sizes ($L$) in the SC lattice. Note the valley of negative value immediately above $T_C$ for $\Delta/J=2.979$.}
\end{figure*}

\begin{figure*}
\includegraphics[width=1.82\columnwidth]{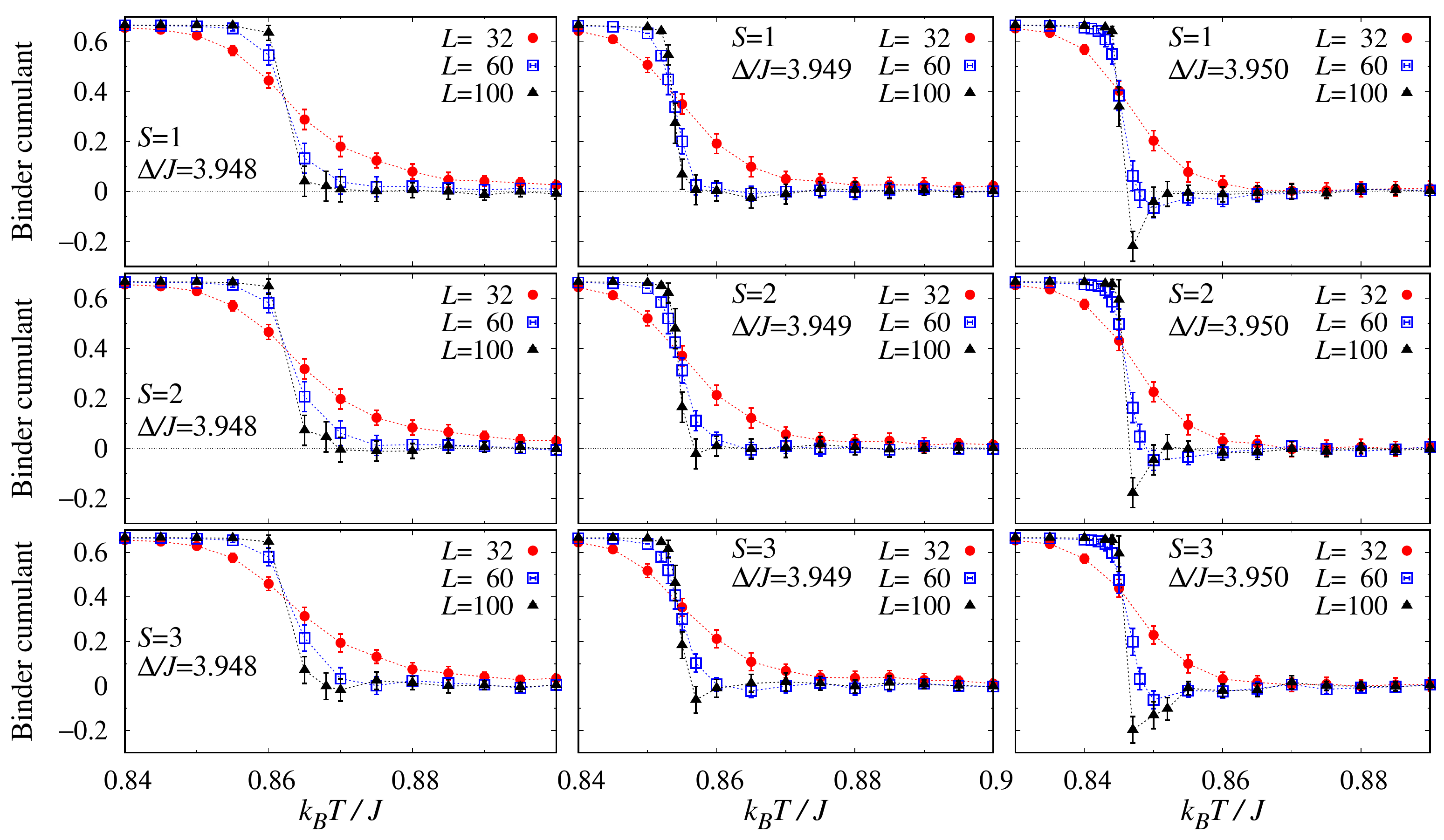}
\caption{\label{fig:fig3}Binder cumulant as a function of temperature ($T$) for various lattice sizes ($L$) in the BCC lattice. Note the valley of negative value immediately above $T_C$ for $\Delta/J=3.950$.}
\end{figure*}

\begin{figure}
\includegraphics[width=0.9\columnwidth]{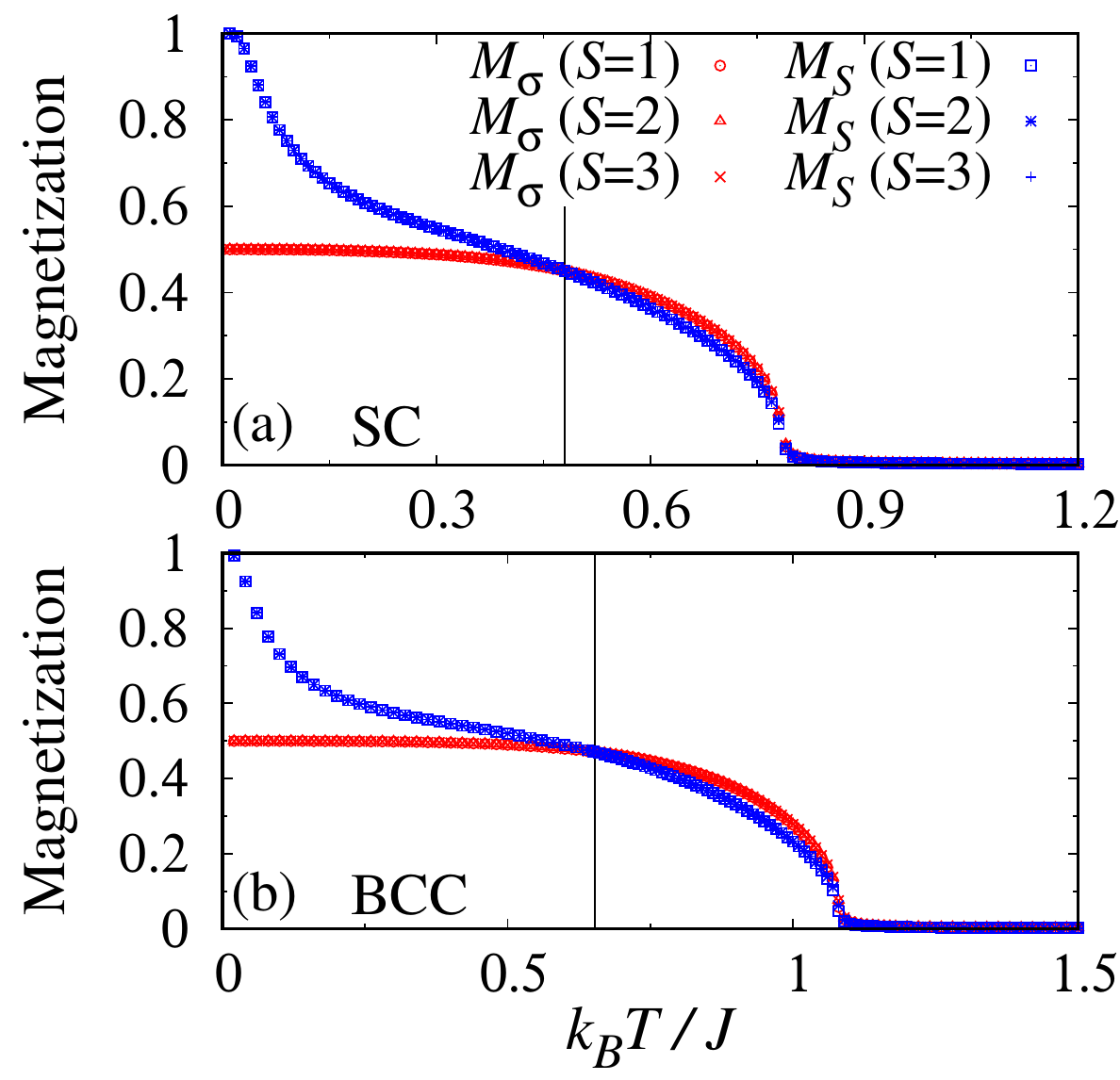}
\caption{\label{fig:fig4} Sublattice magnetizations $M_\sigma$ and $M_S$ as a function of temperature for the SC lattice with $\Delta/J=2.9$ and for the BCC lattice with $\Delta/J=3.9$. 
The linear size of lattices is $L=60$. 
The vertical straight lines indicate the compensation point $T_\mathrm{comp}$ below the critical temperature $T_C$.}
\end{figure}

\begin{figure*}
\includegraphics[width=1.8\columnwidth]{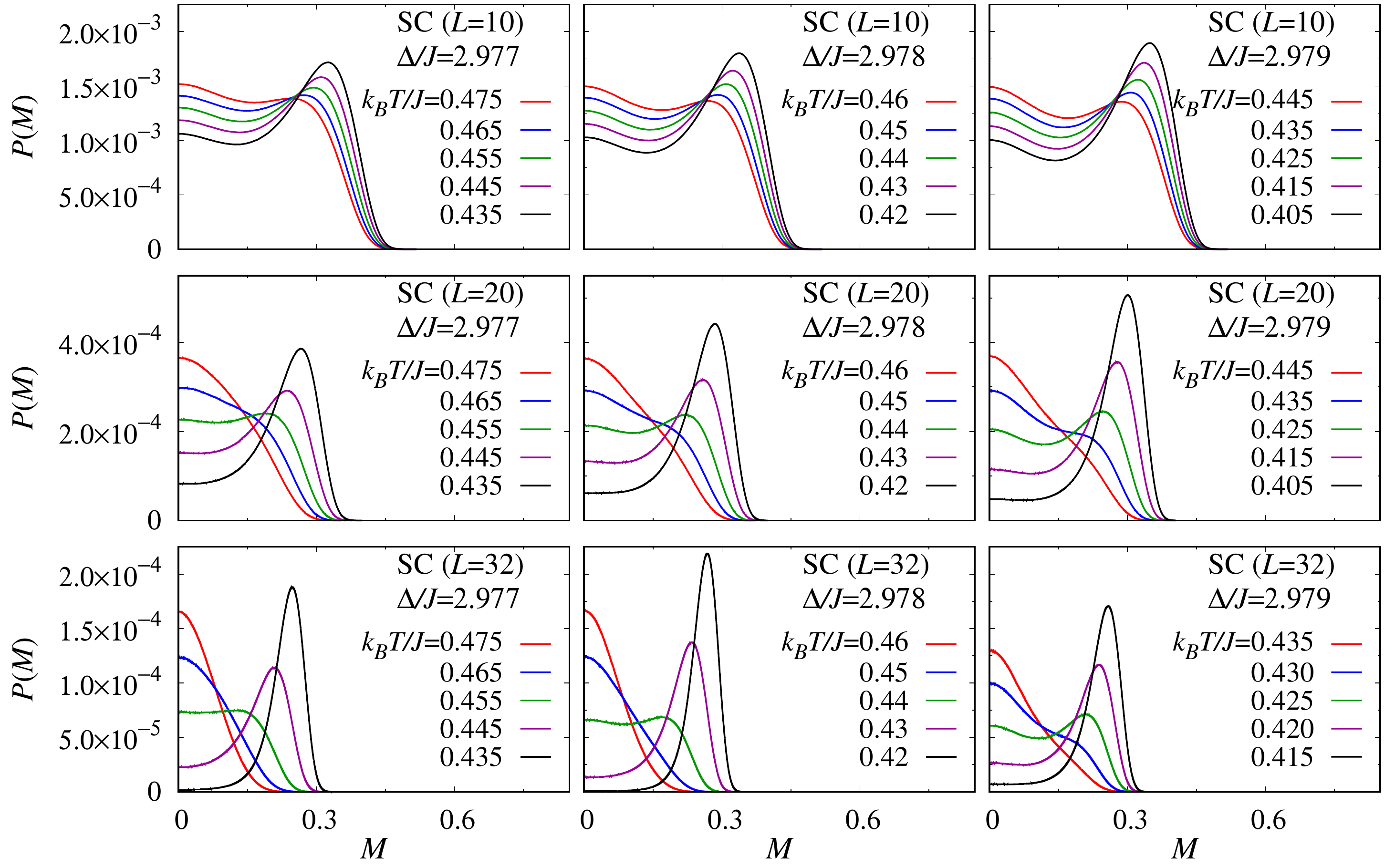}
\caption{\label{fig:fig5}Histogram of the total magnetization $P(M)$ for various values of linear size ($L$) at temperatures crossing the transition in the SC lattice for $S = 1$. $P(M)$ is symmetric about $M = 0$ and only results of positive $M$ are shown.}
\end{figure*}

\begin{figure*}
\includegraphics[width=1.8\columnwidth]{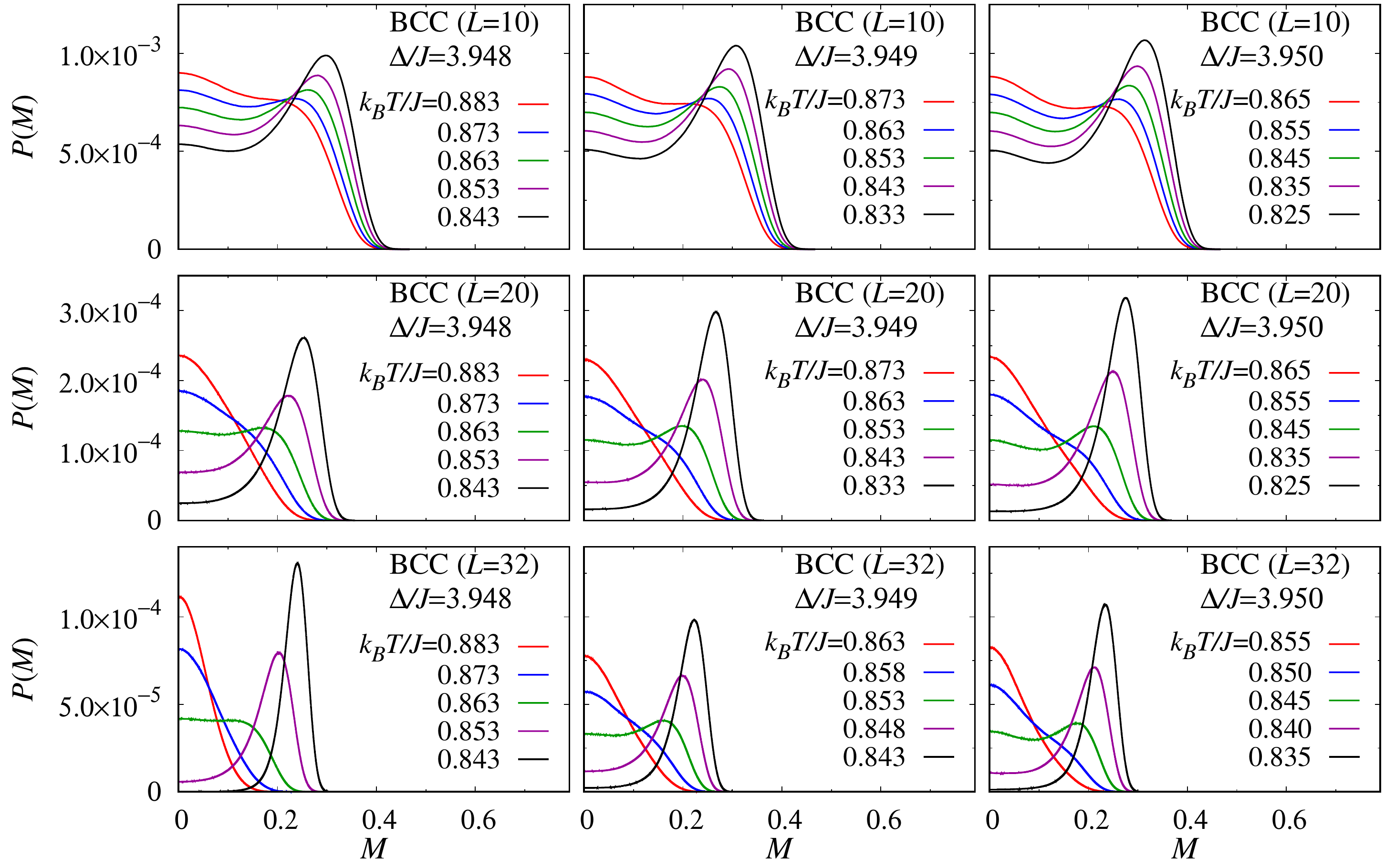}
\caption{\label{fig:fig6}Histogram of the total magnetization $P(M)$ for various values of the lattice size ($L$) at temperatures crossing the transition in the BCC lattice for $S = 1$. $P(M)$ is symmetric about $M = 0$ and only results of positive $M$ are shown.}
\end{figure*}

\begin{figure}
\includegraphics[width=0.9\columnwidth]{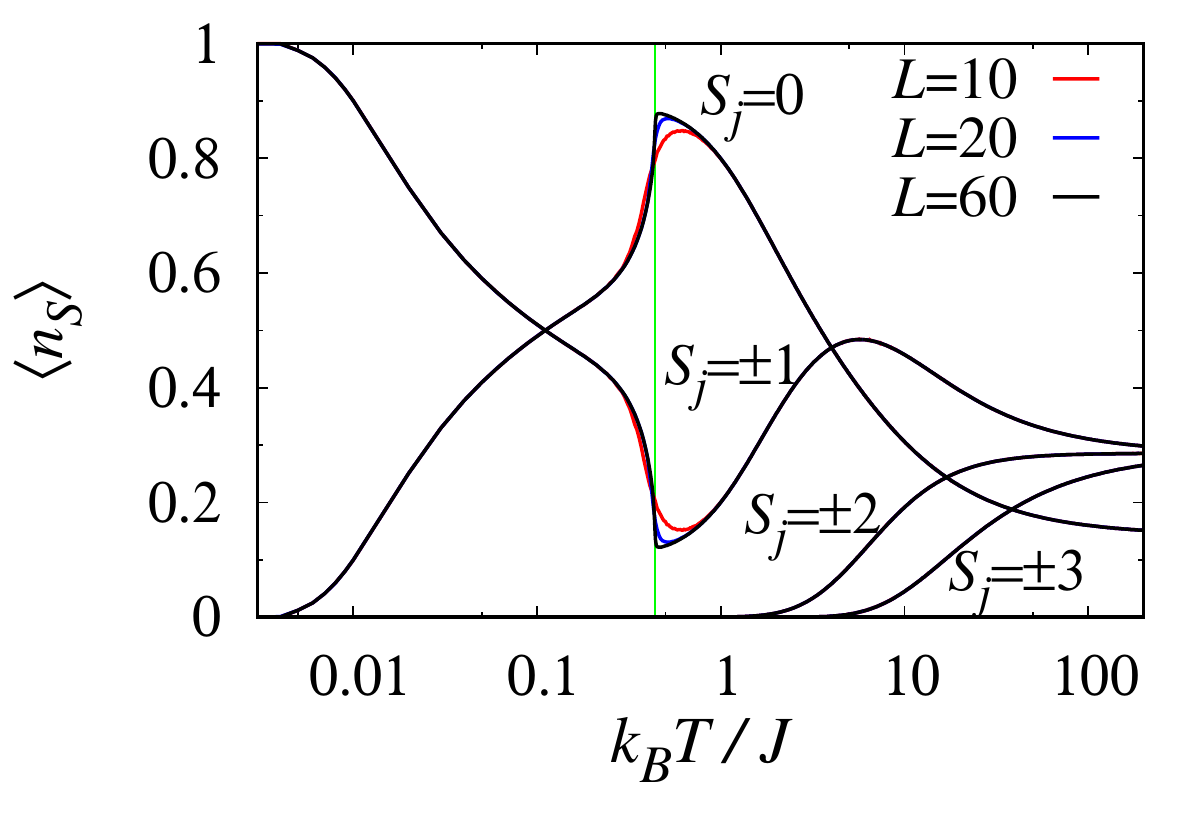}
\caption{\label{fig:fig7}
Density of each spin state $\langle n_S \rangle$ as a function of temperature for $\Delta/J=2.978$ and $S=3$ in the SC lattice. The vertical straight line represents the transition temperature.}
\end{figure}

\begin{figure}
\includegraphics[width=0.9\columnwidth]{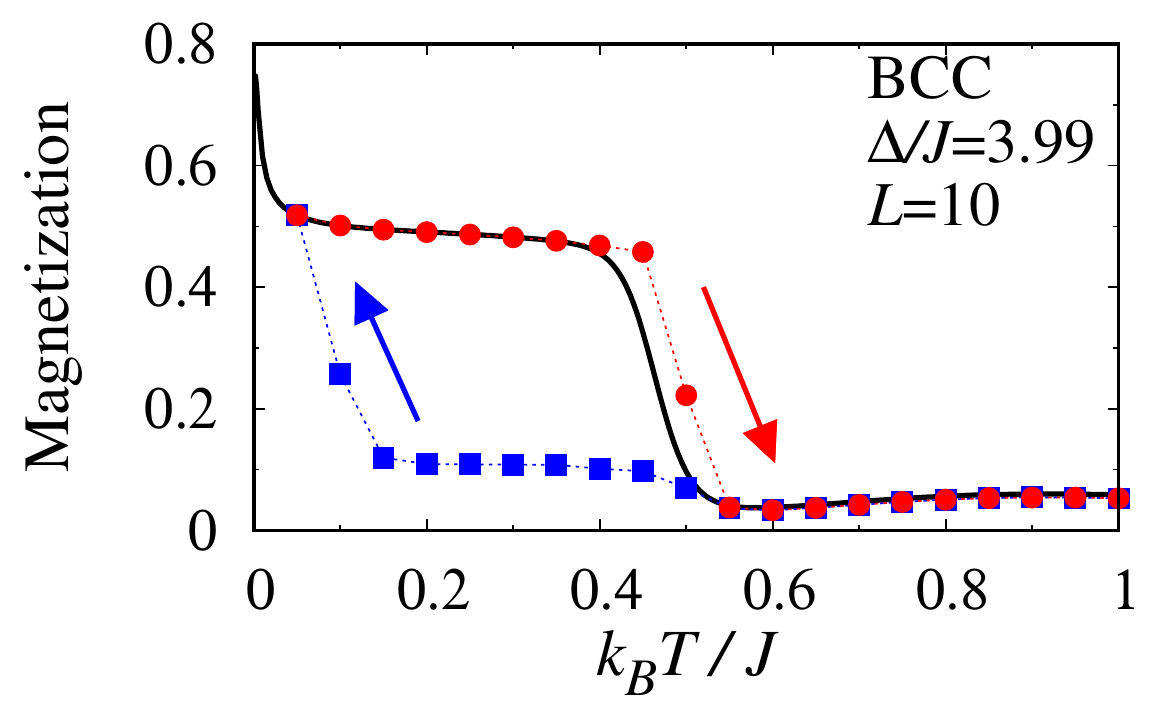}
\caption{\label{fig:fig8}
Total magnetization as a function of temperature in the BCC lattice for $S=1$, $\Delta/J= 3.99$, and $L = 10$ calculated by the MU (symbols). Squares and circles were obtained while cooling-down and warming-up processes, respectively. 
At each temperature, $2\times 10^5$ MC steps were performed; former $10^5$ steps were discarded and only later $10^5$ steps were used to calculate total magnetization. Results of independent 100 runs were averaged for each process.
The results by the WL sampling is plotted by the solid line for comparison.}
\end{figure}

\section{Results and discussion \label{Sec:Results}}

Figure~\ref{fig:fig1} shows phase diagrams in the $T$-$\Delta$ plane for the mixed-spin Blume-Capel model on SC and BCC lattices with $S=1$, $2$, and $3$. 
The critical temperature $T_C$ was obtained by the crossing of Binder cumulant of lattices with different sizes.
This method can be used in first-order as well as continuous phase transitions \cite{Challa86} (see Figs.~\ref{fig:fig2} and \ref{fig:fig3}).
Binder cumulant was calculated by two methods: solid curves and symbols in Fig.~\ref{fig:fig1} represent results from the WL and MU, respectively. 
They are consistent with each other within 1\%. 
For the WL method, the lattice size is limited to $L = 4$, $6$, and $10$ for $S=1$, and $L = 4$ and $6$ for $S>1$; for the MU, much larger lattices ($L = 32$ and $60$) were used. 
We estimate that the error by the correction-to-scaling \cite{Ferrenberg18}, if it exists, is small because the two results by the WL and MU methods are very close to each other. 

In the phase diagram, there are a few qualitatively different regimes according to the value of $\Delta$. For sufficiently large negative crystal field ($\Delta \rightarrow -\infty$), the system undergoes a continuous transition at a nearly constant value of $T_C$ shown in the figures by a horizontal line. 
Because low-spin states ($|S_i| < S$) are suppressed in $\Lambda_2$ sites, the model is reduced into the conventional two-state Ising model with spin-1/2 and spin-$S$ in each sublattice.
We confirmed that the critical temperature converges to $T_C = (S/2)T_C^{\mathrm{Ising}}$ in the limit $\Delta \rightarrow -\infty$ within error bars, which corresponds to the critical temperature of the conventional Ising model: $J/k_B T_C^{\mathrm{Ising}}=0.221654626(5)$ for the SC lattice \cite{Ferrenberg18} and $J/k_B T_C^{\mathrm{Ising}}=0.1573725(6)$ for the BCC lattice \cite{Butera00,Lundow09}.
On the other hand, for $\Delta> \Delta_{\mathrm{crit}}$ the vacancies ($S_j=0$) become dominant and no long-range order occurs in the system since a spin $\sigma_i$ in $\Lambda_1$ is surrounded by vacancies. In fact, the $\Lambda_1$ spins are randomly oriented when the crystal field is greater than $\Delta_{\mathrm{crit}}/J= 3$ and $4$ for the SC and BCC lattices, respectively.

The most interesting part of the phase diagram is the intermediate regime, where the system changes the nature of the transition from continuous to first-order giving rise to a TCP and a line of compensation points. 
As $\Delta$ increases, the critical temperature decreases abruptly because non-zero spin states in $\Lambda_2$ are reduced by the positive crystal field.
As shown in Fig.~\ref{fig:fig4}, the reduction of $M_S$ is strong near $T_C$ and there appears the compensation point $T_\mathrm{comp}$, where $M_\sigma=M_S$, below $T_C$. 
In the ferrimagnetic case ($J<0$), the total magnetization becomes zero at $T_\mathrm{comp}$. We show that the compensation appears in the BCC lattice as well as in the SC lattice (see Figs.~\ref{fig:fig1}(c) and \ref{fig:fig1}(d)).
We found that the compensation point does not depend on the magnitude of spin $S$.
The critical temperature and compensation lines decrease with $\Delta$ and vanish at $\Delta>\Delta_\mathrm{crit}$. 


To find the evidence of the discontinuous nature of the transition and to differentiate the first-order from continuous transitions, we used  two methods. 
First, for the first-order transition the Binder cumulant has a valley of negative values immediately above $T_C$, while in the case of the continuous transition, it monotonically decreases to zero as the temperature increases \cite{Challa86,Vollmayr93}. 
Note that the valley of the Binder cumulant can be missing in small-sized lattices even if larger lattices show it. 
Therefore, the existence of the valley indicates that the transition is of first-order, but its absence does not guarantee that the transition is of continuous. 
Figures~\ref{fig:fig2} and \ref{fig:fig3} show that $\Delta_t/J < 2.979$ and $\Delta_t/J < 3.950$ for the TCP in the SC and BCC lattices, respectively. 
Interestingly, Binder cumulant does not depend on the spin magnitude $S$ around the TCP.
The second method is based on the histogram of the order parameter close to $T_C$; the order parameter refers to the total magnetization $M$ in this case. 
For the first-order transition, the histogram of the order parameter has three peaks at $M=0$ and $M=\pm M_0$ with $M_0>0$ close to $T_C$; the central peak increases as temperature increases.
On the other hand, in the continuous transition, there are only two peaks at $M=\pm M_0$ below $T_C$, and $M_0$ decreases as temperature increases to make only one peak at $M=0$ above $T_C$.
Therefore, the existence of the three-peak structure near $T_C$ is the evidence for the discontinuity of the transition.
This method was used by Selke and Oitmaa to estimate the TCP for the SC lattice with $S=1$ \cite{Selke10}. 
However, as shown in Figs.~\ref{fig:fig5} and \ref{fig:fig6}, there exists a large finite-size effect in this method, too. 
Even when a three-peak structure is observed in small-sized lattices, it could disappear in larger lattices. 
For example, for $\Delta/J=2.977$ in the SC lattice in the left column of Fig.~\ref{fig:fig5}, a three-peak structure is clear in $L=10$ but it disappears for $L=32$. 
Therefore, the three-peak structure does not ensure the first-order nature of the transition, while the missing of the three-peak structure indicates that the transition is indeed continuous. 
As a result, we conclude that $\Delta_t/J > 2.977$ and $\Delta_t/J > 3.948$ for the SC and BCC lattices, respectively. 
Ignoring this effect leads to an underestimation of $\Delta_t/J$. Previous Monte Carlo simulations are limited to small lattice sizes $L=4$, which estimated the value of anisotropy of the TCP to be $\Delta_t/J=2.955(15)$ ~\cite{Selke10}.
Though not shown here, we obtained the same results for the cases of $S=2$ and $S=3$ as the $S=1$ case within error bars.
Combining the two results of the Binder cumulant and order parameter histogram, our final conclusion is that $2.977 < \Delta_t/J < 2.979$ and $3.948 < \Delta_t/J < 3.950$ for the SC and BCC lattices, respectively. 
Therefore, we estimate the tricritical point as ($\Delta_t/J=2.978(1)$; $k_B T_t/J=0.439(1)$) and ($\Delta_t/J=3.949(1)$; $k_B T_t/J=0.854(1)$) for the SC and BCC lattices, respectively. Note that the position of the TCP is independent of the spin magnitude $S$. 

To understand the independence of the TCP on $S$, we examined $\langle n_S \rangle$, which is the portion of spin state $S_j=\pm S$ in sublattice $\Lambda_2$, as a function of temperature.
We concentrate on the case $S=3$.
Figure \ref{fig:fig7} shows $\langle n_S \rangle$ as a function of temperature for $\Delta=\Delta_t$ in the SC lattice with $S=3$. 
The size-dependence is very small except around $T_C$.
At very high temperature above $T_C$, $\langle n_0 \rangle$ approaches $1/7$, and $\langle n_1 \rangle$, $\langle n_2 \rangle$, and $\langle n_3 \rangle$ approach $2/7$, as expected.
As the temperature decreases, high spin states ($|S_j| > 1$) are fully suppressed well-above $T_C$ and so they have no role in the transition at the tricritical point. 
Therefore, it is natural that the cases of $S=2$ and $S=3$ have the same tricritical point as the case of $S=1$. 
Below $T_C$, $\langle n_1 \rangle$ reaches 1 smoothly; 
this behavior is contrary to the the Blume-Capel model, where 
$\langle n_1 \rangle$ jumps abruptly to 1 immediately below $T_C$ \cite{Kwak15}. 
We confirmed the same behavior also around first-order transitions with larger $\Delta$.
The discrepancy may be explained by the existence of two interpenetrating sublattices in our model coupled to each other via the interaction $J$. 
One of these sublattices is occupied by $\sigma_i=\pm 1/2$, which tries to force the spin of the last sublattice to be aligned (ferromagnetic) or anti-aligned (ferrimagnetic).

Finally, as the value of $\Delta$ increases passing through the TCP, continuous phase transition changes into first-order transition. 
At first-order transitions, canonical simulations such as MU may be trapped in a metastable phase giving rise to the supercritical slowing down and hysteresis phenomena.
It becomes more serious as $\Delta$ approaches $\Delta_\mathrm{crit}$, for larger $S$, and in larger lattices. 
The hysteresis also depends on the number of MC steps and the speed of temperature change.
We observed no hysteresis in continuous transitions and close to the tricritical point.
In Fig.~\ref{fig:fig8}, we present the thermal dependence of the total magnetization while increasing and lowering temperature obtained by the MU in the BCC lattice for $\Delta/J=3.99$ and $L=10$, which shows a very strong hysteresis effect even in a relatively small lattice.
More MC steps may reduce the hysteresis effect, but we verified the existence of hysteresis at least up to $2\times10^7$ MC steps per each temperature value in this case.
Therefore, the MU should be used  with special care for $\Delta> \Delta_t$.
The WL method overcomes the supercritical slowing down and hence no hysteresis is observed, demonstrating the effectiveness of the extended ensemble method in the mixed-spin systems.

\section{Conclusions \label{Sec:Conclusion}}

We studied the mixed spin-1/2 and spin-$S$ Blume-Capel model with $S=1$, $2$, and $3$ on three-dimensional lattices (SC and BCC) using the MU and the WL sampling to construct phase diagrams.
Although the WL sampling is restricted to small-sized lattices, 
the results by the two algorithms coincide and the error by the correction-to-scaling is estimated to be small.
In the WL method, thermodynamic quantities at arbitrary temperature and single-site anisotropy $\Delta$ can be obtained by just one calculation and there is no supercritical slowing down.
Therefore, it is now clear that the WL scheme is very efficient to study mixed-spin systems. 
At low values of the anisotropy $\Delta$, the mixed-spin system shows critical lines for each integer $S$, which end in first-order transition lines, and they meet at the TCP ($\Delta_t/J$; $k_B T_t/J$).
From the Binder cumulant and the histogram of magnetization as a function of temperature, we determined the TCP with very high precision as ($\Delta_t/J=2.978(1)$; $k_B T_t/J=0.439(1)$) and ($\Delta_t/J=3.949(1)$; $k_B T_t/J=0.854(1)$) for the SC and BCC lattices, respectively. 
The location of the TCP is independent of $S$ because higher spin states of $|S_j|>1$ are suppressed close to the TCP, which is confirmed by the density of each spin state $\langle n_S \rangle$ as a function of temperature. 
In addition, we demonstrated the existence of the line of compensation points in both lattices, which is also spin-independent.

\section*{Acknowledgments}
This work was supported by GIST Research Institute (GRI) grant funded by the GIST in 2019.

\section*{References}

\bibliography{Ref}

\begin{thebibliography}{40}%
\makeatletter
\providecommand \@ifxundefined [1]{%
 \@ifx{#1\undefined}
}%
\providecommand \@ifnum [1]{%
 \ifnum #1\expandafter \@firstoftwo
 \else \expandafter \@secondoftwo
 \fi
}%
\providecommand \@ifx [1]{%
 \ifx #1\expandafter \@firstoftwo
 \else \expandafter \@secondoftwo
 \fi
}%
\providecommand \natexlab [1]{#1}%
\providecommand \enquote  [1]{``#1''}%
\providecommand \bibnamefont  [1]{#1}%
\providecommand \bibfnamefont [1]{#1}%
\providecommand \citenamefont [1]{#1}%
\providecommand \href@noop [0]{\@secondoftwo}%
\providecommand \href [0]{\begingroup \@sanitize@url \@href}%
\providecommand \@href[1]{\@@startlink{#1}\@@href}%
\providecommand \@@href[1]{\endgroup#1\@@endlink}%
\providecommand \@sanitize@url [0]{\catcode `\\12\catcode `\$12\catcode
  `\&12\catcode `\#12\catcode `\^12\catcode `\_12\catcode `\%12\relax}%
\providecommand \@@startlink[1]{}%
\providecommand \@@endlink[0]{}%
\providecommand \url  [0]{\begingroup\@sanitize@url \@url }%
\providecommand \@url [1]{\endgroup\@href {#1}{\urlprefix }}%
\providecommand \urlprefix  [0]{URL }%
\providecommand \Eprint [0]{\href }%
\providecommand \doibase [0]{http://dx.doi.org/}%
\providecommand \selectlanguage [0]{\@gobble}%
\providecommand \bibinfo  [0]{\@secondoftwo}%
\providecommand \bibfield  [0]{\@secondoftwo}%
\providecommand \translation [1]{[#1]}%
\providecommand \BibitemOpen [0]{}%
\providecommand \bibitemStop [0]{}%
\providecommand \bibitemNoStop [0]{.\EOS\space}%
\providecommand \EOS [0]{\spacefactor3000\relax}%
\providecommand \BibitemShut  [1]{\csname bibitem#1\endcsname}%
\let\auto@bib@innerbib\@empty
\bibitem [{\citenamefont {Ising}(1925)}]{Ising}%
  \BibitemOpen
  \bibfield  {author} {\bibinfo {author} {\bibfnamefont {E.}~\bibnamefont
  {Ising}},\ }\href {\doibase doi:10.1007/BF02980577} {\bibfield  {journal}
  {\bibinfo  {journal} {Z. Phys.}\ }\textbf {\bibinfo {volume} {31}},\ \bibinfo
  {pages} {253–} (\bibinfo {year} {1925})}\BibitemShut {NoStop}%
\bibitem [{\citenamefont {Onsager}(1944)}]{Onsager}%
  \BibitemOpen
  \bibfield  {author} {\bibinfo {author} {\bibfnamefont {L.}~\bibnamefont
  {Onsager}},\ }\href {\doibase 10.1103/PhysRev.65.117} {\bibfield  {journal}
  {\bibinfo  {journal} {Phys. Rev.}\ }\textbf {\bibinfo {volume} {65}},\
  \bibinfo {pages} {117} (\bibinfo {year} {1944})}\BibitemShut {NoStop}%
\bibitem [{\citenamefont {Ferrenberg}\ \emph {et~al.}(2018)\citenamefont
  {Ferrenberg}, \citenamefont {Xu},\ and\ \citenamefont
  {Landau}}]{Ferrenberg18}%
  \BibitemOpen
  \bibfield  {author} {\bibinfo {author} {\bibfnamefont {A.~M.}\ \bibnamefont
  {Ferrenberg}}, \bibinfo {author} {\bibfnamefont {J.}~\bibnamefont {Xu}}, \
  and\ \bibinfo {author} {\bibfnamefont {D.~P.}\ \bibnamefont {Landau}},\
  }\href {\doibase 10.1103/PhysRevE.97.043301} {\bibfield  {journal} {\bibinfo
  {journal} {Phys. Rev. E}\ }\textbf {\bibinfo {volume} {97}},\ \bibinfo
  {pages} {043301} (\bibinfo {year} {2018})}\BibitemShut {NoStop}%
\bibitem [{\citenamefont {Enting}(1979)}]{Enting79}%
  \BibitemOpen
  \bibfield  {author} {\bibinfo {author} {\bibfnamefont {I.~G.}\ \bibnamefont
  {Enting}},\ }\href {\doibase https://doi.org/10.1016/0003-4916(79)90268-9}
  {\bibfield  {journal} {\bibinfo  {journal} {Ann. Phys. (N. Y.)}\ }\textbf
  {\bibinfo {volume} {123}},\ \bibinfo {pages} {141} (\bibinfo {year}
  {1979})}\BibitemShut {NoStop}%
\bibitem [{\citenamefont {Leuenberger}\ and\ \citenamefont
  {Loss}(2001)}]{Leuenberger01}%
  \BibitemOpen
  \bibfield  {author} {\bibinfo {author} {\bibfnamefont {M.~N.}\ \bibnamefont
  {Leuenberger}}\ and\ \bibinfo {author} {\bibfnamefont {D.}~\bibnamefont
  {Loss}},\ }\href {\doibase 10.1038/35071024} {\bibfield  {journal} {\bibinfo
  {journal} {Nature}\ }\textbf {\bibinfo {volume} {410}},\ \bibinfo {pages}
  {789} (\bibinfo {year} {2001})}\BibitemShut {NoStop}%
\bibitem [{\citenamefont {Berman}\ \emph {et~al.}(1994)\citenamefont {Berman},
  \citenamefont {Doolen}, \citenamefont {Holm},\ and\ \citenamefont
  {Tsifrinovich}}]{Berman94}%
  \BibitemOpen
  \bibfield  {author} {\bibinfo {author} {\bibfnamefont {G.~P.}\ \bibnamefont
  {Berman}}, \bibinfo {author} {\bibfnamefont {G.~D.}\ \bibnamefont {Doolen}},
  \bibinfo {author} {\bibfnamefont {D.~D.}\ \bibnamefont {Holm}}, \ and\
  \bibinfo {author} {\bibfnamefont {V.~I.}\ \bibnamefont {Tsifrinovich}},\
  }\href {\doibase 10.1016/0375-9601(94)90537-1} {\bibfield  {journal}
  {\bibinfo  {journal} {Phys. Lett. A}\ }\textbf {\bibinfo {volume} {193}},\
  \bibinfo {pages} {444} (\bibinfo {year} {1994})}\BibitemShut {NoStop}%
\bibitem [{\citenamefont {Blume}(1966)}]{Blume66}%
  \BibitemOpen
  \bibfield  {author} {\bibinfo {author} {\bibfnamefont {M.}~\bibnamefont
  {Blume}},\ }\href {\doibase 10.1103/PhysRev.141.517} {\bibfield  {journal}
  {\bibinfo  {journal} {Phys. Rev.}\ }\textbf {\bibinfo {volume} {141}},\
  \bibinfo {pages} {517} (\bibinfo {year} {1966})}\BibitemShut {NoStop}%
\bibitem [{\citenamefont {Capel}(1966)}]{Capel66}%
  \BibitemOpen
  \bibfield  {author} {\bibinfo {author} {\bibfnamefont {H.~W.}\ \bibnamefont
  {Capel}},\ }\href {\doibase https://doi.org/10.1016/0031-8914(66)90027-9}
  {\bibfield  {journal} {\bibinfo  {journal} {Physica}\ }\textbf {\bibinfo
  {volume} {32}},\ \bibinfo {pages} {966} (\bibinfo {year} {1966})}\BibitemShut
  {NoStop}%
\bibitem [{\citenamefont {Selke}\ and\ \citenamefont
  {Yeomans}(1983)}]{Selke83}%
  \BibitemOpen
  \bibfield  {author} {\bibinfo {author} {\bibfnamefont {W.}~\bibnamefont
  {Selke}}\ and\ \bibinfo {author} {\bibfnamefont {J.}~\bibnamefont
  {Yeomans}},\ }\href {\doibase 10.1088/0305-4470/16/12/024} {\bibfield
  {journal} {\bibinfo  {journal} {J. Phys. A: Math. Gen.}\ }\textbf {\bibinfo
  {volume} {16}},\ \bibinfo {pages} {2789} (\bibinfo {year}
  {1983})}\BibitemShut {NoStop}%
\bibitem [{\citenamefont {Selke}\ \emph {et~al.}(1984)\citenamefont {Selke},
  \citenamefont {Huse},\ and\ \citenamefont {Kroll}}]{Selke84}%
  \BibitemOpen
  \bibfield  {author} {\bibinfo {author} {\bibfnamefont {W.}~\bibnamefont
  {Selke}}, \bibinfo {author} {\bibfnamefont {D.~A.}\ \bibnamefont {Huse}}, \
  and\ \bibinfo {author} {\bibfnamefont {D.~M.}\ \bibnamefont {Kroll}},\ }\href
  {\doibase 10.1088/0305-4470/17/15/019} {\bibfield  {journal} {\bibinfo
  {journal} {J. Phys. A: Math. Gen.}\ }\textbf {\bibinfo {volume} {17}},\
  \bibinfo {pages} {3019} (\bibinfo {year} {1984})}\BibitemShut {NoStop}%
\bibitem [{\citenamefont {Fytas}\ and\ \citenamefont {Selke}(2013)}]{Fytas13}%
  \BibitemOpen
  \bibfield  {author} {\bibinfo {author} {\bibfnamefont {N.~G.}\ \bibnamefont
  {Fytas}}\ and\ \bibinfo {author} {\bibfnamefont {W.}~\bibnamefont {Selke}},\
  }\href {\doibase 10.1140/epjb/e2013-40475-6} {\bibfield  {journal} {\bibinfo
  {journal} {Eur. Phys. J. B}\ }\textbf {\bibinfo {volume} {86}},\ \bibinfo
  {pages} {365} (\bibinfo {year} {2013})}\BibitemShut {NoStop}%
\bibitem [{\citenamefont {Fytas}\ \emph {et~al.}(2019)\citenamefont {Fytas},
  \citenamefont {Mainou}, \citenamefont {Theodorakis},\ and\ \citenamefont
  {Malakis}}]{Fytas19}%
  \BibitemOpen
  \bibfield  {author} {\bibinfo {author} {\bibfnamefont {N.~G.}\ \bibnamefont
  {Fytas}}, \bibinfo {author} {\bibfnamefont {A.}~\bibnamefont {Mainou}},
  \bibinfo {author} {\bibfnamefont {P.~E.}\ \bibnamefont {Theodorakis}}, \ and\
  \bibinfo {author} {\bibfnamefont {A.}~\bibnamefont {Malakis}},\ }\href
  {\doibase 10.1103/PhysRevE.99.012111} {\bibfield  {journal} {\bibinfo
  {journal} {Phys. Rev. E}\ }\textbf {\bibinfo {volume} {99}},\ \bibinfo
  {pages} {012111} (\bibinfo {year} {2019})}\BibitemShut {NoStop}%
\bibitem [{\citenamefont {Beale}(1986)}]{Beale86}%
  \BibitemOpen
  \bibfield  {author} {\bibinfo {author} {\bibfnamefont {P.~D.}\ \bibnamefont
  {Beale}},\ }\href {\doibase 10.1103/PhysRevB.33.1717} {\bibfield  {journal}
  {\bibinfo  {journal} {Phys. Rev. B}\ }\textbf {\bibinfo {volume} {33}},\
  \bibinfo {pages} {1717} (\bibinfo {year} {1986})}\BibitemShut {NoStop}%
\bibitem [{\citenamefont {Deserno}(1997)}]{Deserno97}%
  \BibitemOpen
  \bibfield  {author} {\bibinfo {author} {\bibfnamefont {M.}~\bibnamefont
  {Deserno}},\ }\href {\doibase 10.1103/PhysRevE.56.5204} {\bibfield  {journal}
  {\bibinfo  {journal} {Phys. Rev. E}\ }\textbf {\bibinfo {volume} {56}},\
  \bibinfo {pages} {5204} (\bibinfo {year} {1997})}\BibitemShut {NoStop}%
\bibitem [{\citenamefont {Silva}\ \emph {et~al.}(2006)\citenamefont {Silva},
  \citenamefont {Caparica},\ and\ \citenamefont {Plascak}}]{Silva06}%
  \BibitemOpen
  \bibfield  {author} {\bibinfo {author} {\bibfnamefont {C.~J.}\ \bibnamefont
  {Silva}}, \bibinfo {author} {\bibfnamefont {A.~A.}\ \bibnamefont {Caparica}},
  \ and\ \bibinfo {author} {\bibfnamefont {J.~A.}\ \bibnamefont {Plascak}},\
  }\href {\doibase 10.1103/PhysRevE.73.036702} {\bibfield  {journal} {\bibinfo
  {journal} {Phys. Rev. E}\ }\textbf {\bibinfo {volume} {73}},\ \bibinfo
  {pages} {036702} (\bibinfo {year} {2006})}\BibitemShut {NoStop}%
\bibitem [{\citenamefont {Fytas}(2011)}]{Fytas11}%
  \BibitemOpen
  \bibfield  {author} {\bibinfo {author} {\bibfnamefont {N.~G.}\ \bibnamefont
  {Fytas}},\ }\href {\doibase 10.1140/epjb/e2010-10738-y} {\bibfield  {journal}
  {\bibinfo  {journal} {Eur. Phys. J. B}\ }\textbf {\bibinfo {volume} {79}},\
  \bibinfo {pages} {21} (\bibinfo {year} {2011})}\BibitemShut {NoStop}%
\bibitem [{\citenamefont {Kwak}\ \emph {et~al.}(2015)\citenamefont {Kwak},
  \citenamefont {Jeong}, \citenamefont {Lee},\ and\ \citenamefont
  {Kim}}]{Kwak15}%
  \BibitemOpen
  \bibfield  {author} {\bibinfo {author} {\bibfnamefont {W.}~\bibnamefont
  {Kwak}}, \bibinfo {author} {\bibfnamefont {J.}~\bibnamefont {Jeong}},
  \bibinfo {author} {\bibfnamefont {J.}~\bibnamefont {Lee}}, \ and\ \bibinfo
  {author} {\bibfnamefont {D.-H.}\ \bibnamefont {Kim}},\ }\href {\doibase
  10.1103/PhysRevE.92.022134} {\bibfield  {journal} {\bibinfo  {journal} {Phys.
  Rev. E}\ }\textbf {\bibinfo {volume} {92}},\ \bibinfo {pages} {022134}
  (\bibinfo {year} {2015})}\BibitemShut {NoStop}%
\bibitem [{\citenamefont {Jung}\ and\ \citenamefont {Kim}(2017)}]{Jung17}%
  \BibitemOpen
  \bibfield  {author} {\bibinfo {author} {\bibfnamefont {M.}~\bibnamefont
  {Jung}}\ and\ \bibinfo {author} {\bibfnamefont {D.-H.}\ \bibnamefont {Kim}},\
  }\href {\doibase 10.1140/epjb/e2017-80471-2} {\bibfield  {journal} {\bibinfo
  {journal} {Eur. Phys. J. B}\ }\textbf {\bibinfo {volume} {90}},\ \bibinfo
  {pages} {245} (\bibinfo {year} {2017})}\BibitemShut {NoStop}%
\bibitem [{\citenamefont {Butera}\ and\ \citenamefont
  {Pernici}(2018)}]{Butera18}%
  \BibitemOpen
  \bibfield  {author} {\bibinfo {author} {\bibfnamefont {P.}~\bibnamefont
  {Butera}}\ and\ \bibinfo {author} {\bibfnamefont {M.}~\bibnamefont
  {Pernici}},\ }\href {\doibase 10.1016/j.physa.2018.05.010} {\bibfield
  {journal} {\bibinfo  {journal} {Physica A}\ }\textbf {\bibinfo {volume}
  {507}},\ \bibinfo {pages} {22} (\bibinfo {year} {2018})}\BibitemShut
  {NoStop}%
\bibitem [{\citenamefont {Drillon}\ \emph {et~al.}(1983)\citenamefont
  {Drillon}, \citenamefont {Coronado}, \citenamefont {Beltran},\ and\
  \citenamefont {Georges}}]{Drillion83}%
  \BibitemOpen
  \bibfield  {author} {\bibinfo {author} {\bibfnamefont {M.}~\bibnamefont
  {Drillon}}, \bibinfo {author} {\bibfnamefont {E.}~\bibnamefont {Coronado}},
  \bibinfo {author} {\bibfnamefont {D.}~\bibnamefont {Beltran}}, \ and\
  \bibinfo {author} {\bibfnamefont {R.}~\bibnamefont {Georges}},\ }\href
  {\doibase 10.1016/0301-0104(83)85267-7} {\bibfield  {journal} {\bibinfo
  {journal} {Chem. Phys.}\ }\textbf {\bibinfo {volume} {79}},\ \bibinfo {pages}
  {449} (\bibinfo {year} {1983})}\BibitemShut {NoStop}%
\bibitem [{\citenamefont {Mathoni\`{e}re}\ \emph {et~al.}(1996)\citenamefont
  {Mathoni\`{e}re}, \citenamefont {Nuttall}, \citenamefont {Carling},\ and\
  \citenamefont {Day}}]{Mathoniere96}%
  \BibitemOpen
  \bibfield  {author} {\bibinfo {author} {\bibfnamefont {C.}~\bibnamefont
  {Mathoni\`{e}re}}, \bibinfo {author} {\bibfnamefont {C.~J.}\ \bibnamefont
  {Nuttall}}, \bibinfo {author} {\bibfnamefont {S.~G.}\ \bibnamefont
  {Carling}}, \ and\ \bibinfo {author} {\bibfnamefont {P.}~\bibnamefont
  {Day}},\ }\href {\doibase 10.1021/ic950703v} {\bibfield  {journal} {\bibinfo
  {journal} {Inorg. Chem.}\ }\textbf {\bibinfo {volume} {35}},\ \bibinfo
  {pages} {1201} (\bibinfo {year} {1996})}\BibitemShut {NoStop}%
\bibitem [{\citenamefont {Gon{\c{c}}alves}(1985)}]{Gon_alves85}%
  \BibitemOpen
  \bibfield  {author} {\bibinfo {author} {\bibfnamefont {L.~L.}\ \bibnamefont
  {Gon{\c{c}}alves}},\ }\href {\doibase 10.1088/0031-8949/32/3/012} {\bibfield
  {journal} {\bibinfo  {journal} {Phys. Scr.}\ }\textbf {\bibinfo {volume}
  {32}},\ \bibinfo {pages} {248} (\bibinfo {year} {1985})}\BibitemShut
  {NoStop}%
\bibitem [{\citenamefont {Dakhama}(1998)}]{Dakhama98}%
  \BibitemOpen
  \bibfield  {author} {\bibinfo {author} {\bibfnamefont {A.}~\bibnamefont
  {Dakhama}},\ }\href {\doibase 10.1016/S0378-4371(97)00583-9} {\bibfield
  {journal} {\bibinfo  {journal} {Physica A}\ }\textbf {\bibinfo {volume}
  {252}},\ \bibinfo {pages} {225} (\bibinfo {year} {1998})}\BibitemShut
  {NoStop}%
\bibitem [{\citenamefont {Dakhama}\ \emph {et~al.}(2018)\citenamefont
  {Dakhama}, \citenamefont {Azhari},\ and\ \citenamefont
  {Benayad}}]{Dakhama18}%
  \BibitemOpen
  \bibfield  {author} {\bibinfo {author} {\bibfnamefont {A.}~\bibnamefont
  {Dakhama}}, \bibinfo {author} {\bibfnamefont {M.}~\bibnamefont {Azhari}}, \
  and\ \bibinfo {author} {\bibfnamefont {N.}~\bibnamefont {Benayad}},\ }\href
  {\doibase 10.1088/2399-6528/aacbbe} {\bibfield  {journal} {\bibinfo
  {journal} {J. Phys. Commun.}\ }\textbf {\bibinfo {volume} {2}},\ \bibinfo
  {pages} {065011} (\bibinfo {year} {2018})}\BibitemShut {NoStop}%
\bibitem [{\citenamefont {Zhang}\ and\ \citenamefont {Yang}(1993)}]{Zhang93}%
  \BibitemOpen
  \bibfield  {author} {\bibinfo {author} {\bibfnamefont {G.-M.}\ \bibnamefont
  {Zhang}}\ and\ \bibinfo {author} {\bibfnamefont {C.-Z.}\ \bibnamefont
  {Yang}},\ }\href {\doibase 10.1103/PhysRevB.48.9452} {\bibfield  {journal}
  {\bibinfo  {journal} {Phys. Rev. B}\ }\textbf {\bibinfo {volume} {48}},\
  \bibinfo {pages} {9452} (\bibinfo {year} {1993})}\BibitemShut {NoStop}%
\bibitem [{\citenamefont {Buend{\'{\i}}a}\ and\ \citenamefont
  {Novotny}(1997)}]{Buendia97}%
  \BibitemOpen
  \bibfield  {author} {\bibinfo {author} {\bibfnamefont {G.~M.}\ \bibnamefont
  {Buend{\'{\i}}a}}\ and\ \bibinfo {author} {\bibfnamefont {M.~A.}\
  \bibnamefont {Novotny}},\ }\href {\doibase 10.1088/0953-8984/9/27/021}
  {\bibfield  {journal} {\bibinfo  {journal} {J. Phys.: Condens. Matter}\
  }\textbf {\bibinfo {volume} {9}},\ \bibinfo {pages} {5951} (\bibinfo {year}
  {1997})}\BibitemShut {NoStop}%
\bibitem [{\citenamefont {Buend{\'{\i}}a}\ and\ \citenamefont
  {Cardona}(1999)}]{Buendia99}%
  \BibitemOpen
  \bibfield  {author} {\bibinfo {author} {\bibfnamefont {G.~M.}\ \bibnamefont
  {Buend{\'{\i}}a}}\ and\ \bibinfo {author} {\bibfnamefont {R.}~\bibnamefont
  {Cardona}},\ }\href {\doibase 10.1103/PhysRevB.59.6784} {\bibfield  {journal}
  {\bibinfo  {journal} {Phys. Rev. B}\ }\textbf {\bibinfo {volume} {59}},\
  \bibinfo {pages} {6784} (\bibinfo {year} {1999})}\BibitemShut {NoStop}%
\bibitem [{\citenamefont {Selke}\ and\ \citenamefont {Oitmaa}(2010)}]{Selke10}%
  \BibitemOpen
  \bibfield  {author} {\bibinfo {author} {\bibfnamefont {W.}~\bibnamefont
  {Selke}}\ and\ \bibinfo {author} {\bibfnamefont {J.}~\bibnamefont {Oitmaa}},\
  }\href {\doibase 10.1088/0953-8984/22/7/076004} {\bibfield  {journal}
  {\bibinfo  {journal} {J. Phys. Condens. Matter}\ }\textbf {\bibinfo {volume}
  {22}},\ \bibinfo {pages} {076004} (\bibinfo {year} {2010})}\BibitemShut
  {NoStop}%
\bibitem [{\citenamefont {Benayad}(1990)}]{Benayad90}%
  \BibitemOpen
  \bibfield  {author} {\bibinfo {author} {\bibfnamefont {N.}~\bibnamefont
  {Benayad}},\ }\href {\doibase 10.1007/BF01454220} {\bibfield  {journal}
  {\bibinfo  {journal} {Z. Phys. B}\ }\textbf {\bibinfo {volume} {81}},\
  \bibinfo {pages} {99} (\bibinfo {year} {1990})}\BibitemShut {NoStop}%
\bibitem [{\citenamefont {Metropolis}\ \emph {et~al.}(1953)\citenamefont
  {Metropolis}, \citenamefont {Rosenbluth}, \citenamefont {Rosenbluth},
  \citenamefont {Teller},\ and\ \citenamefont {Teller}}]{Metropolis53}%
  \BibitemOpen
  \bibfield  {author} {\bibinfo {author} {\bibfnamefont {N.}~\bibnamefont
  {Metropolis}}, \bibinfo {author} {\bibfnamefont {A.~W.}\ \bibnamefont
  {Rosenbluth}}, \bibinfo {author} {\bibfnamefont {M.~N.}\ \bibnamefont
  {Rosenbluth}}, \bibinfo {author} {\bibfnamefont {A.~H.}\ \bibnamefont
  {Teller}}, \ and\ \bibinfo {author} {\bibfnamefont {E.}~\bibnamefont
  {Teller}},\ }\href {\doibase 10.1063/1.1699114} {\bibfield  {journal}
  {\bibinfo  {journal} {J. Chem. Phys.}\ }\textbf {\bibinfo {volume} {21}},\
  \bibinfo {pages} {1087} (\bibinfo {year} {1953})}\BibitemShut {NoStop}%
\bibitem [{\citenamefont {Quadros}\ and\ \citenamefont
  {Salinas}(1994)}]{Quadros94}%
  \BibitemOpen
  \bibfield  {author} {\bibinfo {author} {\bibfnamefont {S.~G.~A.}\
  \bibnamefont {Quadros}}\ and\ \bibinfo {author} {\bibfnamefont {S.~R.}\
  \bibnamefont {Salinas}},\ }\href {\doibase 10.1016/0378-4371(94)90319-0}
  {\bibfield  {journal} {\bibinfo  {journal} {Physica A}\ }\textbf {\bibinfo
  {volume} {206}},\ \bibinfo {pages} {479} (\bibinfo {year}
  {1994})}\BibitemShut {NoStop}%
\bibitem [{\citenamefont {Wang}\ and\ \citenamefont {Landau}(2001)}]{Wang01}%
  \BibitemOpen
  \bibfield  {author} {\bibinfo {author} {\bibfnamefont {F.}~\bibnamefont
  {Wang}}\ and\ \bibinfo {author} {\bibfnamefont {D.~P.}\ \bibnamefont
  {Landau}},\ }\href {\doibase 10.1103/PhysRevLett.86.2050} {\bibfield
  {journal} {\bibinfo  {journal} {Phys. Rev. Lett.}\ }\textbf {\bibinfo
  {volume} {86}},\ \bibinfo {pages} {2050} (\bibinfo {year}
  {2001})}\BibitemShut {NoStop}%
\bibitem [{\citenamefont {Kumar}\ and\ \citenamefont {Yusuf}(2015)}]{Kumar15}%
  \BibitemOpen
  \bibfield  {author} {\bibinfo {author} {\bibfnamefont {A.}~\bibnamefont
  {Kumar}}\ and\ \bibinfo {author} {\bibfnamefont {S.}~\bibnamefont {Yusuf}},\
  }\href {\doibase 10.1016/j.physrep.2014.10.003} {\bibfield  {journal}
  {\bibinfo  {journal} {Phys. Rep.}\ }\textbf {\bibinfo {volume} {556}},\
  \bibinfo {pages} {1} (\bibinfo {year} {2015})}\BibitemShut {NoStop}%
\bibitem [{\citenamefont {Janke}(1994)}]{Janke94}%
  \BibitemOpen
  \bibfield  {author} {\bibinfo {author} {\bibfnamefont {W.}~\bibnamefont
  {Janke}},\ }in\ \href@noop {} {\emph {\bibinfo {booktitle} {Computer
  Simulation Studies in Condensed-Matler Physics {VII}}}},\ \bibinfo {editor}
  {edited by\ \bibinfo {editor} {\bibfnamefont {D.~P.}\ \bibnamefont {Landau}},
  \bibinfo {editor} {\bibfnamefont {K.~K.}\ \bibnamefont {Mon}}, \ and\
  \bibinfo {editor} {\bibfnamefont {H.-B.}\ \bibnamefont {Sch\"{u}ttler}}}\
  (\bibinfo  {publisher} {Springer},\ \bibinfo {address} {Berlin},\ \bibinfo
  {year} {1994})\ p.~\bibinfo {pages} {29}\BibitemShut {NoStop}%
\bibitem [{\citenamefont {Yu}(2015)}]{Yu15}%
  \BibitemOpen
  \bibfield  {author} {\bibinfo {author} {\bibfnamefont {U.}~\bibnamefont
  {Yu}},\ }\href {\doibase 10.1103/PhysRevE.91.062121} {\bibfield  {journal}
  {\bibinfo  {journal} {Phys. Rev. E}\ }\textbf {\bibinfo {volume} {91}},\
  \bibinfo {pages} {062121} (\bibinfo {year} {2015})}\BibitemShut {NoStop}%
\bibitem [{\citenamefont {Binder}(1981)}]{Binder81}%
  \BibitemOpen
  \bibfield  {author} {\bibinfo {author} {\bibfnamefont {K.}~\bibnamefont
  {Binder}},\ }\href {\doibase 10.1007/BF01293604} {\bibfield  {journal}
  {\bibinfo  {journal} {Z. Phys. B}\ }\textbf {\bibinfo {volume} {43}},\
  \bibinfo {pages} {119} (\bibinfo {year} {1981})}\BibitemShut {NoStop}%
\bibitem [{\citenamefont {Challa}\ \emph {et~al.}(1986)\citenamefont {Challa},
  \citenamefont {Landau},\ and\ \citenamefont {Binder}}]{Challa86}%
  \BibitemOpen
  \bibfield  {author} {\bibinfo {author} {\bibfnamefont {M.~S.~S.}\
  \bibnamefont {Challa}}, \bibinfo {author} {\bibfnamefont {D.~P.}\
  \bibnamefont {Landau}}, \ and\ \bibinfo {author} {\bibfnamefont
  {K.}~\bibnamefont {Binder}},\ }\href {\doibase 10.1103/PhysRevB.34.1841}
  {\bibfield  {journal} {\bibinfo  {journal} {Phys. Rev. B}\ }\textbf {\bibinfo
  {volume} {34}},\ \bibinfo {pages} {1841} (\bibinfo {year}
  {1986})}\BibitemShut {NoStop}%
\bibitem [{\citenamefont {Butera}\ and\ \citenamefont {Comi}(2000)}]{Butera00}%
  \BibitemOpen
  \bibfield  {author} {\bibinfo {author} {\bibfnamefont {P.}~\bibnamefont
  {Butera}}\ and\ \bibinfo {author} {\bibfnamefont {M.}~\bibnamefont {Comi}},\
  }\href {\doibase 10.1103/PhysRevB.62.14837} {\bibfield  {journal} {\bibinfo
  {journal} {Phys. Rev. B}\ }\textbf {\bibinfo {volume} {62}},\ \bibinfo
  {pages} {14837} (\bibinfo {year} {2000})}\BibitemShut {NoStop}%
\bibitem [{\citenamefont {Lundow}\ \emph {et~al.}(2009)\citenamefont {Lundow},
  \citenamefont {Markstr{\"{o}}m},\ and\ \citenamefont {Rosengren}}]{Lundow09}%
  \BibitemOpen
  \bibfield  {author} {\bibinfo {author} {\bibfnamefont {P.~H.}\ \bibnamefont
  {Lundow}}, \bibinfo {author} {\bibfnamefont {K.}~\bibnamefont
  {Markstr{\"{o}}m}}, \ and\ \bibinfo {author} {\bibfnamefont {A.}~\bibnamefont
  {Rosengren}},\ }\href {\doibase 10.1080/14786430802680512} {\bibfield
  {journal} {\bibinfo  {journal} {Phil. Mag.}\ }\textbf {\bibinfo {volume}
  {89}},\ \bibinfo {pages} {2009} (\bibinfo {year} {2009})}\BibitemShut
  {NoStop}%
\bibitem [{\citenamefont {Vollmayr}\ \emph {et~al.}(1993)\citenamefont
  {Vollmayr}, \citenamefont {Reger}, \citenamefont {Scheucher},\ and\
  \citenamefont {Binder}}]{Vollmayr93}%
  \BibitemOpen
  \bibfield  {author} {\bibinfo {author} {\bibfnamefont {K.}~\bibnamefont
  {Vollmayr}}, \bibinfo {author} {\bibfnamefont {J.~D.}\ \bibnamefont {Reger}},
  \bibinfo {author} {\bibfnamefont {M.}~\bibnamefont {Scheucher}}, \ and\
  \bibinfo {author} {\bibfnamefont {K.}~\bibnamefont {Binder}},\ }\href
  {\doibase 10.1007/BF01316713} {\bibfield  {journal} {\bibinfo  {journal} {Z.
  Phys. B}\ }\textbf {\bibinfo {volume} {91}},\ \bibinfo {pages} {113}
  (\bibinfo {year} {1993})}\BibitemShut {NoStop}%
\end{thebibliography}%

\end{document}